\documentclass[
superscriptaddress,
reprint,
 amsmath,amssymb,
 aps,
 prl,
]{revtex4-1}

\usepackage{graphicx}
\usepackage{dcolumn}
\usepackage{bm}
\usepackage{amsmath}
\usepackage{amsfonts}
\usepackage{amssymb}
\usepackage{dsfont}
\usepackage{color}
\usepackage{xcolor}
\usepackage{siunitx}
\usepackage[colorlinks=true,
    linkcolor=blue,
    filecolor=blue,      
    urlcolor=blue,
    citecolor = blue]{hyperref}
\usepackage{braket}
\usepackage{mathtools}
\usepackage[mathlines]{lineno}
\usepackage[normalem]{ulem}
\usepackage{scalerel}
\usepackage{ bbold }
\usepackage{inputenc}
\usepackage[T1]{fontenc}

\graphicspath{{./figures/}}

\begin{document}

\title{Non-Abelian gauge invariance from dynamical decoupling} 

\date{\today}

\author{Valentin Kasper}
\affiliation{ICFO - Institut de Ciencies Fotoniques, The Barcelona Institute of Science and Technology, Av. Carl Friedrich Gauss 3, 08860 Castelldefels (Barcelona), Spain}
\affiliation{Department of Physics, Harvard University, Cambridge, MA, 02138, USA}
\email{valentin.kasper@icfo.eu}
\author{Torsten V. Zache}
\affiliation{Heidelberg University, Institut für Theoretische Physik, Philosophenweg 16, 69120 Heidelberg, Germany}
\affiliation{Center for Quantum Physics, University of Innsbruck, 6020 Innsbruck, Austria}
\affiliation{Institute for Quantum Optics and Quantum Information of the Austrian Academy of Sciences, 6020 Innsbruck, Austria}
\author{Fred Jendrzejewski}
\affiliation{Universit\"at  Heidelberg,  Kirchhoff-Institut  f\"ur  Physik, Im  Neuenheimer  Feld  227,  69120  Heidelberg,  Germany }
\author{Maciej Lewenstein}
\affiliation{ICFO - Institut de Ciencies Fotoniques, The Barcelona Institute of Science and Technology, Av. Carl Friedrich Gauss 3, 08860 Castelldefels (Barcelona), Spain}
\affiliation{ICREA, Pg. Lluis Companys 23, 08010 Barcelona, Spain}
\author{Erez Zohar}
\affiliation{Racah Institute of Physics, The Hebrew University of Jerusalem, Givat Ram, Jerusalem 91904, Israel}
\begin{abstract}
Lattice gauge theories are fundamental to such distinct fields as particle physics, condensed matter or quantum information theory. The recent progress in the control of artificial quantum systems already allows for studying Abelian lattice gauge theories in table-top experiments. However, the realization of non-Abelian models remains challenging.
Here, we employ a coherent quantum control scheme to enforce non-Abelian gauge invariance, and discuss this idea in detail for a one dimensional SU(2) lattice gauge system.
We comment on how to extend our scheme to other non-Abelian gauge symmetries and higher spatial dimensions. Because of its wide applicability, the  presented coherent control scheme provides a promising route for the quantum simulation of non-Abelian lattice gauge theories.
\end{abstract}
\maketitle

\emph{Introduction.} Gauge theories are of central importance for such different fields as particle physics~\cite{Weinberg2005}, condensed matter~\cite{Fradkin2013}, and quantum information theory~\cite{Kitaev2003}. They are crucial to understand quantum chromodynamics~\cite{Gross1973, Aoki2017}, are used to describe anyonic excitations~\cite{Kitaev2006}, but are also applied to model electronic transport in strongly correlated materials~\cite{Lee2009}. While gauge theories are postulated to describe fundamental interactions~\cite{Yang1954}, they emerge as an effective description at low energies~\cite{Moessner2001}, but one can also imagine artificial quantum systems with a local symmetry~\cite{Wiese2013, Zohar2016, Dalmonte2016, Banuls2020}. 
Recent experimental progress made the idea of quantum simulating Abelian lattice gauge theories a reality, which can now be accessed with trapped ions~\cite{Martinez2016, Kokail2019}, superconducting qubits~\cite{Klco2018} or ultracold atom systems~\cite{Bernien2017,Gorg2019,Schweizer2019,Mil2020, Yang2020}. 

Whereas the experimental progress for Abelian lattice gauge theories is remarkable, the quantum simulation of non-Abelian lattice gauge theories remains challenging.
Several theoretical works proposed already the quantum simulation of non-Abelian lattice gauge theories~\cite{Zohar2013, Banerjee2013, Tagliacozzo2013, Stannigel2014}. 
However, their implementation is difficult, because one has to fine tune atomic collisions, employ intricate energy penalties, use elaborate digital schemes, or engineer dissipation to enforce non-Abelian gauge invariance. Consequently, the design of feasible quantum simulators for non-Abelian lattice
gauge theories receives considerable attention~\cite{Raychowdhury2020, Klco2020, Kasper2020, Davoudi2020, Dasgupta2020}.

In this work we present how to implement non-Abelian lattice gauge theories 
using techniques from quantum control theory~\cite{DAlessandro2008}, known as dynamical decoupling~\cite{Viola1999}. In particular, we propose a periodic protocol, whose net effect is to enforce a local symmetry. 
Subsequently, we demonstrate the dynamical decoupling approach by numerical calculations. We conclude by commenting on its applicability 
to a wide range of platforms, and its generalization to higher spatial dimensions, different gauge groups and formulations of lattice gauge theories~\cite{Horn1981, Mathur2007, Zohar2015b}.



\begin{figure}[t!]
    \centering
    \includegraphics{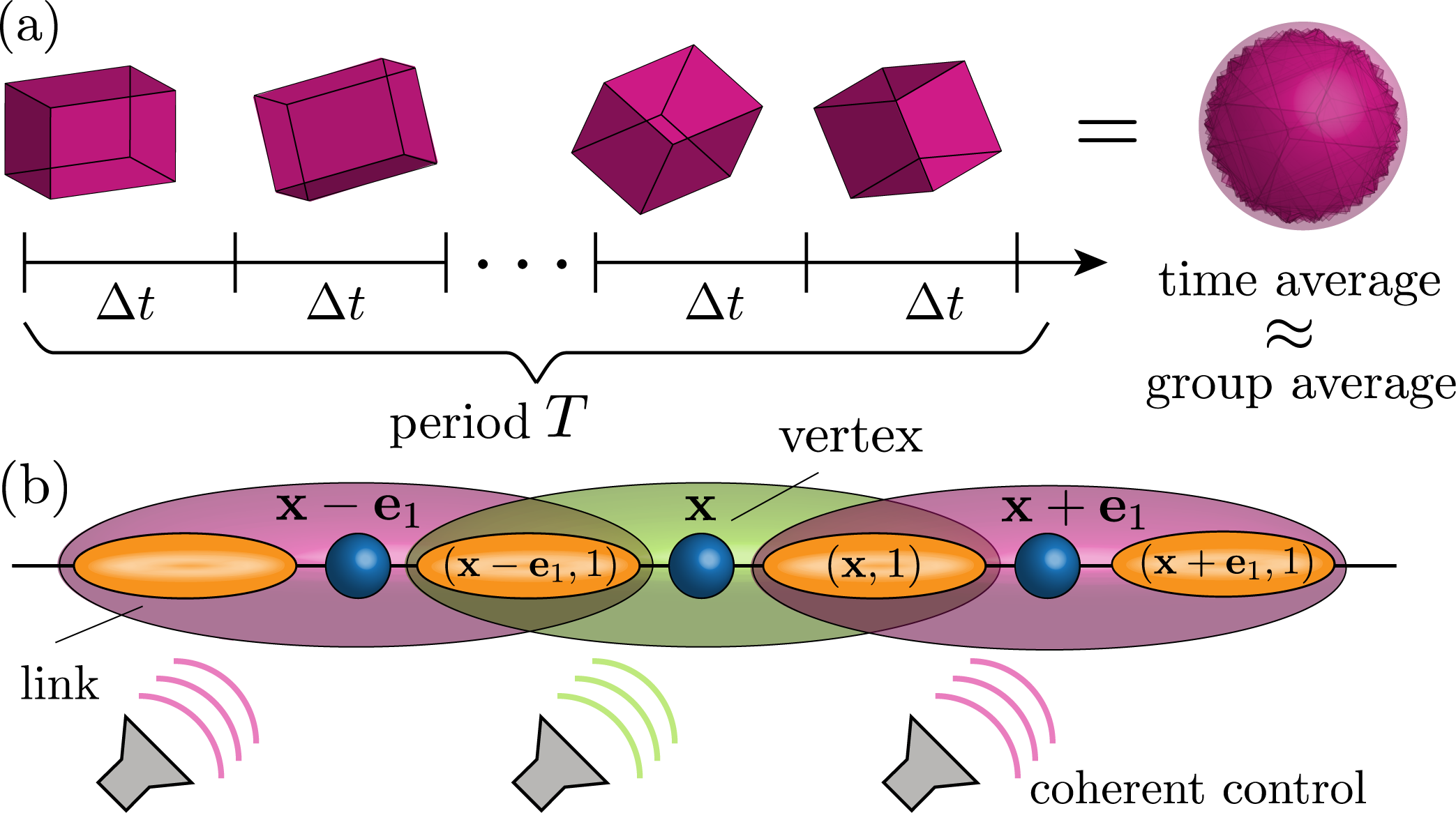}
    \caption{\textbf{Dynamical decoupling for quantum simulating non-Abelian lattice gauge theories} (a) Temporal sequence of local non-Abelian symmetry transformations visualized by rotating a rigid body over a period $T$. The non-Abelian symmetry transformation is applied for a time interval $\Delta t$ and then changed to the next transformation. The time average over this sequence of unitary transformations leads to an effective symmetry, such that the time average, see Eq.~\eqref{eq:TimeAveraging}, corresponds to a group average (b) Coherent control applied to a lattice system enforces as local symmetry.
    Purple and green areas illustrate a local, staggered coherent drive, which renders the system gauge invariant after time averaging. The necessary locality constraints on the Hamiltonian are explained in the main text. For the one dimensional example considered here fermions $\psi_m(\mathbf{x})$ reside on the vertices, while a single boson $b_{lm}(\mathbf{x},i)$ resides on the link $(\mathbf{x},i)$.}
    
    \label{fig:Overview}
\end{figure}


\emph{Local symmetry from periodic driving. }
The key characteristic of lattice gauge theories is the presence of a local symmetry. This local 
 symmetry is generated by the set of local operators $G_{a}(\mathbf{x})$,    
where $a$ enumerates the operators and $\mathbf{x}$ denotes the lattice site. Further, the $G_{a}(\mathbf{x})$ commute with the lattice gauge theory Hamiltonian $H_{\text{LGT}}$ for all $\mathbf{x}$ and $a$. In artifical quantum systems such local symmetries are typically not present, but can be approximately realized.
Recent proposals used periodically driven atoms~\cite{Barbiero2019}, atomic mixtures~\cite{Zohar2013b, Kasper2017}, or digital approaches~\cite{Muschik2017,Klco2018}
to realize Abelian lattice gauge theories.
Among these ideas, periodic driving~\cite{Lin2011, Bakr2011, Struck2012, Miyake2013, Jotzu2014}
appears promising, because it has been already used for static non-Abelian gauge fields~\cite{Hauke2012, Goldman2014b}. It is an immediate question, whether one can
implement a quantum simulator of dynamical non-Abelian gauge fields via a periodic drive. 
{Related questions have been discussed in~\cite{Stannigel2014} and more recently in~\cite{ Lamm2020, Tran2021,Halimeh2020c, Halimeh2020}.}

{Experimental realizations of lattice gauge Hamiltonians 
$H_{\text{LGT}}$ with quantum simulators may experience perturbing contributions $H_{\text{P}}$ that violate gauge invariance~\cite{Halimeh2020, Halimeh2020c, halimeh2020fate}}. We propose to add a coherent drive $H_1(t)$ with period $T$ to the static contribution $H_0 = H_{\text{LGT}} + H_{\text{P}}$ to suppress such perturbing terms,
see Fig.~\ref{fig:Overview}. 
For a strong, coherent, and periodic drive $H_1(t)$, the time evolution is determined by an effective Hamiltonian
\begin{align}
    \bar{H}_0 \equiv   \frac{1}{T} \int_{0}^{T} d t \, U_1^{\dagger}(t) H_0 U_1(t)  \label{eq:Average}\, ,
\end{align}
which is given by the first order of the Magnus expansion~\cite{Goldman2014, Bukov2015, Eckardt2017} in the interaction picture with the unitary $U_{1}(t) \equiv \mathcal{T} e^{ -i \int_{0}^{t} d u H_{1}(u) }$ and time ordering operator $\mathcal{T}$. The question is how to choose the drive $H_1$ to generate a gauge invariant effective Hamiltonian $\bar{H}_0$. 
In the following, we provide an explicit example of a lattice gauge theory and the construction of $H_1$ that enforces a local SU(2) symmetry.

\emph{Lattice gauge Hamiltonian.}\!
{For concreteness we start with a SU(2) gauge invariant lattice Hamiltonian in $d$ space dimensions~\cite{Kogut1975} and work with staggered fermions~\cite{Susskind1977}. There are two fermionic modes per vertex $\mathbf{x}$ with corresponding field operators $\psi_m(\mathbf{x})$ and $m \in \{\uparrow, \downarrow\}$. For staggered fermions~\cite{Susskind1977} the purely fermionic part of the Hamiltonian is
		$H_{{F}}=M \sum_{\mathbf{x},m}(-1)^{s(\mathbf{x})} \psi_{m}^{\dagger}(\mathbf{x}) \psi_{m}(\mathbf{x})$,
where $s(\mathbf{x}) = x_1+ \ldots x_d $. The total number of fermions equals $N_L$, the number of lattice sites, also called half-filling for two modes per site. 
The local fermionic charge is given by ${Q}_a(\mathbf{x})=\frac{1}{2} \sum_{kl} \psi_{k}^{\dagger}(\mathbf{x}) \sigma^a_{kl} \psi_{l}(\mathbf{x})$, which fulfills the SU(2) Lie algebra $\left[Q_{a}(\mathbf{x}), Q_{b}(\mathbf{y})\right]=i \varepsilon_{a b c} Q_{c}(\mathbf{x}) \delta_{\mathbf{x},\mathbf{y}} $ where the $\sigma_a$ denote the Pauli-matrices.}

{The gauge  degrees of freedom are rigid bodies and reside on the links~\cite{Kogut1975}. The right and left electric field operators are $R_a$ and $L_a$ respectively $(a=1,2,3)$, satisfy the Lie algebras $\left[R_{a}, R_{b}\right] =i \varepsilon_{a b c} R_{c} $, $\left[L_{a}, L_{b}\right] =-i \varepsilon_{a b c} L_{c}$, $\left[L_{a}, R_{b}\right] =0$, with the Casimir operator $\mathbf{J}^{2}\equiv \sum_{a} R_{a} R_{a}=  \sum_{a} L_{a} L_{a} $. The link Hilbert space can be spanned by the eigenstates of the maximal set of mutually commuting operators: $\left|jmn\right\rangle$ such that $\mathbf{J}^2\left|jmn\right\rangle=j\left(j+1\right)\left|jmn\right\rangle$, $L_3\left|jmn\right\rangle=m\left|jmn\right\rangle$ and $R_3\left|jmn\right\rangle=n\left|jmn\right\rangle$.
The interaction of the matter and the gauge field is given by
	\begin{equation}
		H_{{I}}=\epsilon \sum_{\mathbf{x}, i}\left[ \psi_{m}^{\dagger}(\mathbf{x}) U_{m n}(\mathbf{x}, i) \psi_{n}(\mathbf{x}+\mathbf{e}_i)+ \text{H.c.} \right],
	\end{equation}
where $\mathbf{e}_i$ is a unit vector in the $i$ direction and $\epsilon$ is the coupling between matter and gauge field. For the matrix elements and algebra of  $U_{mn}$, see~\footnote{See Supplemental Material.}.
The pure gauge Hamiltonian is given by $H_{G}=H_E+H_B$, with the electric part $H_{{E}}= \frac{g^2}{2} \sum_{\mathbf{x}, i} \mathbf{J}^{2}(\mathbf{x}, i)$ and the magnetic part $H_B= -\frac{1}{g^2} \sum_{\mathbf{x}, i<j} \text{Tr}
	[U(\mathbf{x},i)
	U(\mathbf{x}+\mathbf{e}_i,j)
	U^{\dagger}(\mathbf{x}+\mathbf{e}_j,i)
	U^{\dagger}(\mathbf{x},j)]
+ \text{H.c.} ]$, where $g$ is the gauge coupling constant.}

For our purposes the lattice gauge Hamiltonian is the Kogut-Susskind Hamiltonian $H_{\text{LGT}}= H_G + H_M + H_I$ possesses a SU(2) gauge symmetry, i.e., there exist operators $G_{a}(\mathbf{x})$ such that $\left[G_{a}(\mathbf{x}), G_b(\mathbf{y})\right]= -i \delta_{\mathbf{x},\mathbf{y}} \sum_c \varepsilon_{a b c} G_{c}(\mathbf{x}) $ and  $[H_{\textrm{LGT}},G_{a}(\mathbf{x})]=0$ for all $\mathbf{x}$ and $a$.
For the Kogut-Susskind Hamiltonian the generators, also known as the Gauss's law operators, are
\begin{align}
	G_{a}(\mathbf{x})= \sum_{i} \left[ L_{a}(\mathbf{x}, i)- R_{a}(\mathbf{x}-\mathbf{e}_i, i)\right] -Q_{a}(\mathbf{x}) \,.
\end{align}
Knowing the generators of the local symmetry allows us to design a coherent drive $H_1(t)$, 
which suppresses gauge invariance violating terms as we show in the following.

\emph{Designing the drive.}
The enforcing of a global symmetry through periodic driving was studied in the context of quantum control theory, where a famous solution is known as dynamical decoupling~\cite{Viola1999, Lidar2013}.
Here, we will employ dynamical decoupling to impose local symmetries.
First, we focus on a single vertex $\mathbf{x}$ with the drive $H_1(t) = \sum_{a} \lambda_{a,\mathbf{x}}(t) G_a(\mathbf{x})$ 
and tunable coefficients $\lambda_{a,\mathbf{x}}(t)$. {For example, in an ultracold atom experiment the $\lambda_{a,\mathbf{x}}(t)$ can be implemented via 
Raman transitions between the internal states of the fermionic and bosonic atoms forming $L_a$, $R_a$ and $Q_a$.}

By choosing an appropriate drive $\lambda_{a,\mathbf{x}}(t)$ one samples all gauge transformations $V_{\mathbf{x}}$ and hence may render the effective Hamiltonian $\bar{H}_0$ in Eq.~\eqref{eq:Average} 
gauge invariant at position
$\mathbf{x}$.
In particular, the drive $H_1(t)$ allows one to generate the unitary
\begin{align}
	V_{\mathbf{x}}(g) \equiv e^{-i \alpha G_{3}(\mathbf{x})} e^{-i \beta G_{2}(\mathbf{x})} e^{-i \gamma G_{3}(\mathbf{x})}
	\, , \label{eq:GlobalSU2} \
\end{align}
which is a local SU(2) transformation~\cite{Tung1985}  at $\mathbf{x}$ with the Euler angles $\alpha \in[0, 2\pi)$, $\beta \in[0, \pi ]$, and $\gamma \in[0 ,4 \pi)$.

This sampling of all gauge transformations can be formalized via group averaging,
which is at the heart of many dynamical decoupling schemes. Group averaging $\Pi_{\mathbf{x}}(\cdot)$ symmetrizes an operator $\mathcal{O}$ with respect to a group, and for our quantum simulation scenario we particularly aim at symmetrizing the Hamiltonian $H_0$ with respect to  gauge transformations. The group averaging is given by  
\begin{align}
	\Pi_{\mathbf{x}}(\mathcal{O}) = \frac{1}{\text{Vol}(G)} \int d\mu(g) V^{\dagger}_{\mathbf{x}}(g) \mathcal{O} V_{\mathbf{x}}(g) \, ,
	\label{eq:GroupAveraging}
\end{align}
where $\int d\mu(g)$ is the Haar measure, which for SU(2) is
\begin{align}
	\frac{1}{\text{Vol}(G)} \! \int \! d\mu(g) =  \frac{1}{16\pi^2}\int_{0}^{2 \pi}\!\!\! \mathrm{d} \alpha \int_{-1}^{1}\!\!\! \mathrm{d}(\cos \beta) \int_{0}^{4 \pi} \!\!\! \mathrm{d} \gamma
	\, .
\end{align}
The group averaging in Eq.~\eqref{eq:GroupAveraging} is constructed such that  $\Pi_{\mathbf{x}}(\mathcal{O})$ commutes with every group element of SU(2) as we prove in the supplementary material~\footnote{See Supplemental Material at [URL will be inserted by publisher] for a proof of the group averaging property in the continuous case.}.


In order to find an explicit drive $\lambda_{a,\mathbf{x}}(t)$, that realizes Eq.~\eqref{eq:Average}, we approximate the integral of Eq.~\eqref{eq:GroupAveraging} by a finite sum
\begin{align}
	\Pi_{\mathbf{x}}(\mathcal{O}) \approx \frac{1}{N^3} \sum_{\boldsymbol{\mu}} V^{\dagger}_{\mathbf{x}}(g_{\boldsymbol{\mu}}) \mathcal{O} V_{\mathbf{x}}(g_{\boldsymbol{\mu}}) \, , 
\end{align}
where we chose a three dimensional Cartesian grid to discretize the cube $[0, 2\pi) \times [0, \pi ] \times [0 ,4 \pi)$. The lattice points 
are denoted by $\boldsymbol{\mu}=(\mu_1,\mu_2,\mu_3)$ with $0\leq \mu_i \leq N -1$ and $N$ 
being a non-negative integer. By mapping~\cite{Knuth2005} the multi-index $\boldsymbol{\mu}$ on a linear index $\nu$ and interpreting the linear index as a label for a discretized time step, the group averaging can be written as a one dimensional integral over time
\begin{align}
	\Pi_{\mathbf{x}}(\mathcal{O}) \approx \frac{1}{T} \int_{0}^{T} d t \, U_1^{\dagger}(t) {\mathcal{O}} U_1(t) \,  \label{eq:TimeAveraging}
\end{align}
with the piecewise time evolution
\begin{align}
	U_1(t) \equiv V_{\mathbf{x}}(g_\nu) \quad \text{for } \nu \Delta t \leq t<(\nu+1) \Delta t \, 
\end{align}
and the discretized time step $\Delta t = T/N^3$.
The above recipe of implementing
group averaging via time averaging can be also employed for finite groups and compact Lie groups, such as the dihedral group or SU(N).
 
 {{
Eq.~\eqref{eq:TimeAveraging} represents the leading order in a Magnus expansion~\cite{Goldman2014}, which is a systematic, and therefore controllable high-frequency expansion regularly used  in quantum simulator experiments~\cite{Struck2012, Struck2013,Aidelsburger2013, Jotzu2014, Tai2017, Gorg2019}. The leading order contribution can be made dominant by increasing the frequency of the drive accordingly. The range of validity of the Magnus expansion is model dependent, but can be precisely estimated by calculating higher orders. For example, the high frequency expansion was used to engineer a minimal instance of a $Z_2$ lattice gauge theory~\cite{Schweizer2019}, where the effect of the sub-leading order scales as ${J^{2}}/{(\hbar \omega)}$ with the tunnel coupling $J$ and the driving frequency $\omega$}.}

The generalization of group averaging on each vertex $\mathbf{x}$ is given by
\begin{align}
	\Pi(\mathcal{O}) = \frac{1}{\text{Vol}(G)^{N_L}} \prod_{\mathbf{x}} \int d\mu(g_{\mathbf{x}}) V^{\dagger}_{\mathbf{x}}(g_{\mathbf{x}}) \mathcal{O} V_{\mathbf{x}}(g_{\mathbf{x}}) \, ,
\end{align}
where $N_L$ is the number of lattice sites. 
Repeating the argument of discretizing the group manifold on each lattice site 
and interpreting the group averaging process as a time evolution allows one to enforce gauge invariance on the total system with a periodic drive $H_1(t) = \sum_{a,\mathbf{x}} \lambda_{a,\mathbf{x}}(t) G_{a}(\mathbf{x})$.

\begin{figure}
	\centering
	\includegraphics[width = \columnwidth]{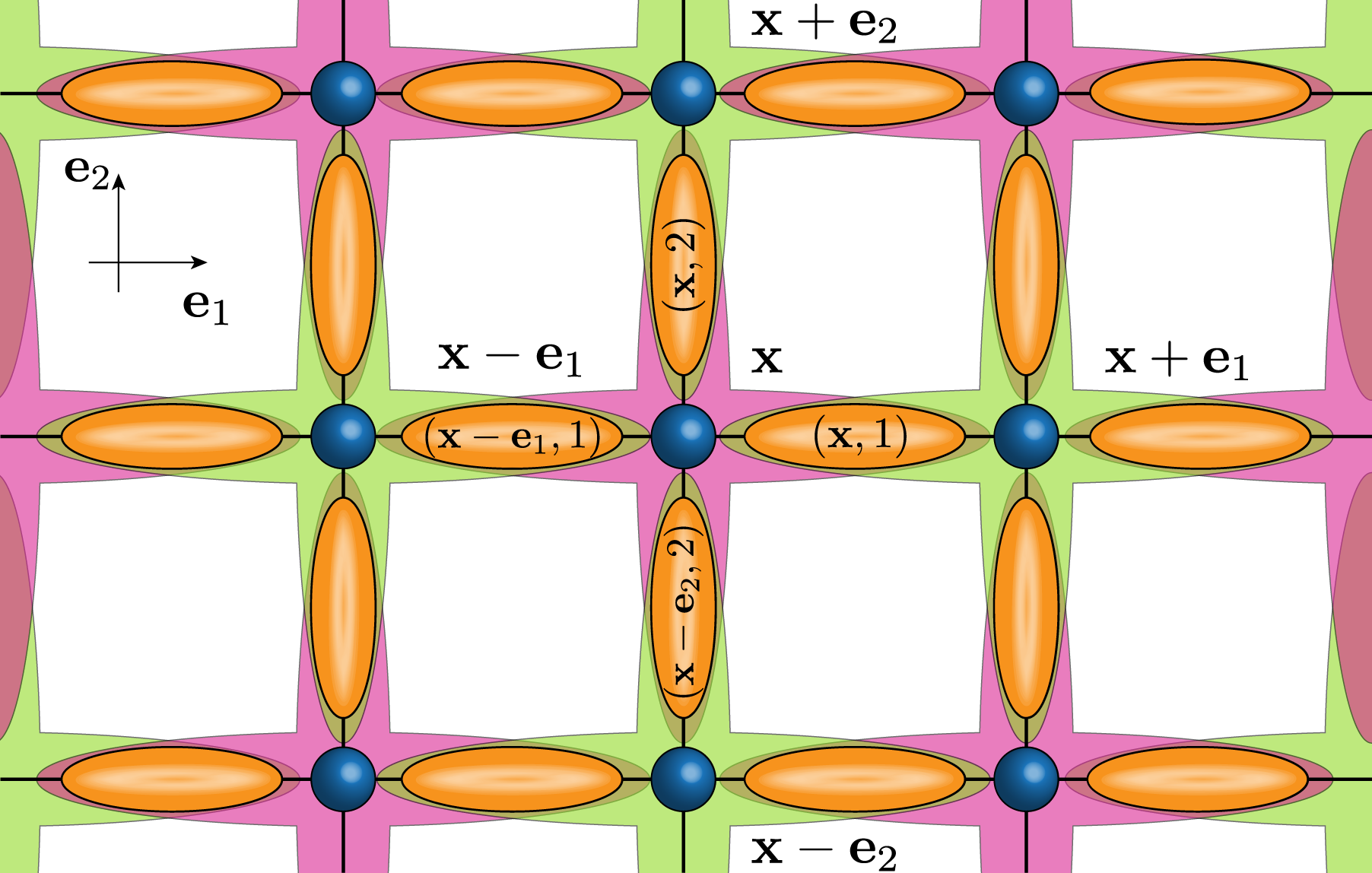}
	\caption{\textbf{Generalizations of local dynamical decoupling scheme:} Fermions on the vertices and bosons on the links. {Considering a Hamiltonian only with local terms and interactions between neighboring field operators, two independent global drives are sufficient to ensure local gauge invariance. The green and purple areas indicate two coherent, but global drives. } }
	\label{fig:HigherDimensions}
\end{figure}

The sampling of $N_L$ Haar integrals is inefficient for $N_L \gg 1 $, but can be drastically reduced, when the Hamiltonian $H_0$ is a sum of local operators. 
We observe, that one only has to perform group averaging at $\mathbf{x}$, if the local contribution of $H_0$ and $V_\mathbf{x}$ do not commute. For the important situation, where the single particle wavefunction on the vertex only overlaps with the single particle wavefunction on the neighboring links, the number of independent Haar integrals can be reduced to two, see Fig.~\ref{fig:HigherDimensions}. This situation is particularly relevant for the majority of previous analog quantum simulator proposals.

\emph{Numerical demonstration.} 
{The dynamical decoupling scheme relies on the algebra of the Gauss's law operators. We employ
the algebra and symmetry preserving truncation schemes of~\cite{Zohar2015b}, where the smallest non-trivial truncation enforces a Hilbert space dimension of five. This Hilbert space is realized with one boson in five modes per link,
$b_{ml}(\mathbf{x},i)$ with $(ml) \in \{00,\uparrow \uparrow, \downarrow \downarrow, \uparrow \downarrow, \downarrow \uparrow\}$  corresponding to the singlet $\left|000\right\rangle$, and the four $\left|\frac{1}{2}mn\right\rangle$ states. The left, right electric field and link variable 
are given by ${L}_a\equiv\frac{1}{2} \sum_{lmn} b_{m l}^{\dagger} \sigma^a_{nm} b_{n l}$ and ${R}_a\equiv\frac{1}{2} \sum_{lmn} b_{l m}^{\dagger} \sigma^a_{m n} b_{l n}$, and
\begin{align}
	U_{mn}(\mathbf{x}, i)=\frac{1}{\sqrt{2}}\left(\begin{array}{cc}
		b_{\uparrow \uparrow}^{\dagger} b_{00}+b_{00}^{\dagger} b_{\downarrow \downarrow} & b_{\uparrow \downarrow}^{\dagger} b_{00}-b_{00}^{\dagger} b_{\downarrow \uparrow} \\
		b_{\downarrow \uparrow}^{\dagger} b_{00}-b_{00}^{\dagger} b_{\uparrow \downarrow} & b_{00}^{\dagger} b_{\uparrow \uparrow}+b_{\downarrow \downarrow}^{\dagger} b_{00}
	\end{array}\right) \, .
\end{align}
These operators satisfy the same commutation relations as in the full Kogut-Susskind case~\cite{Zohar2015b}. Other truncation schemes, which change the symmetry group are possible, but may lead to fake phase transitions~\cite{Collaboration2019}. We further focus on the one dimensional case, where $H_{\text{LGT}} = H_F + H_I + H_E$, to illustrate the dynamical decoupling scheme.}

\begin{figure}
    \centering
    \includegraphics[width = \columnwidth]{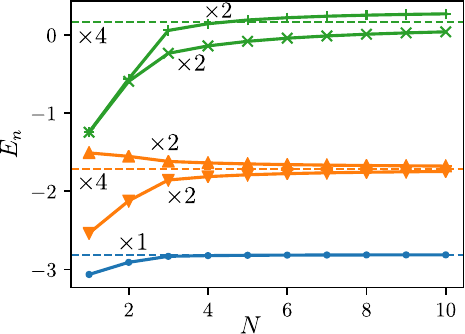}
    \caption{\textbf{Convergence of eigenvalues:} Consider the perturbation $H_{P} =  \sum_{\mathbf{x},a} \gamma_a Q_a(\mathbf{x})$ where $(\gamma_1, \gamma_2, \gamma_3)/M = (0.5, 1.5, 3.5)$ parametrizes the asymmetry of the local charge. This perturbation violates the gauge invariance because of $[G_a, Q_b] = -i\sum_b\varepsilon_{abc} Q_c$. We consider $N_L = 2$ lattice sites and $N$ denotes the number of discretization points of the SU(2) manifold. Depicted are the eigenvalues and degeneracies $E_n$ of $H= H_{\rm{LGT}} + H^N_{{P}}$ for $g^2/(2M) = 0.9 $, $\epsilon/M= 1.1 $, $H^N_{{P}}$ denotes the group averaging of $H_{P}$ with discretization $N$.  The multiplicities next to the lines indicate the degeneracy of the 
    eigenvalue.
    Enforcing the Gauss's law $G_a(\mathbf{x}) \ket{\psi} = 0$ on the eigenstates would further reduce the degeneracy. For $N = 10$ we observe convergence on the percent-level for the first three lowest eigenvalues.}
    \label{fig:Numerics}
\end{figure}

In order to quantify the dynamical decoupling approach we study the convergence of the low
energy eigenvalues of $\bar{H}_0$ towards the low energy eigenvalues of ${H}_{\rm{LGT}}$ with respect to the number of discretization points $N$.
As an explicit example, we consider the following fermionic perturbation $H_{P} =  \sum_{\mathbf{x},a} \gamma_a Q_a(\mathbf{x})$ where $\gamma_a$ parametrizes the asymmetry of the local charge term. In Fig.~\ref{fig:Numerics} we plot the eigenvalues $E_n$ for given $M$, $g$, $\epsilon$, and $\gamma_a$ as a function of $N$ and observe convergence towards the Kogut-Susskind
eigenvalues (dashed lines) already for a moderate discretization of the SU(2) manifold ($N_L = 2$). 
We stress that the convergence properties depend on the perturbation, 
as well as on the concrete dynamical decoupling scheme.

As another example we consider direct tunneling as a perturbation, i.e.,
$H_P = -t \sum_{\mathbf{x},i} \left[ \psi^{\dagger}_m(\mathbf{x})\psi_m(\mathbf{x}+\mathbf{e}_i) + \text{H.c.} \right]$ with the tunneling amplitude $t$. 
The tunneling term can be very efficiently supressed via our dynamical decoupling scheme. 
The group averaging in Eq.~\eqref{eq:Average}  corresponds to three successive averages first with respect $G_3(\mathbf{x})$, then $G_2(\mathbf{x})$
and finally $G_3(\mathbf{x})$ again. Performing the continuous integral with respect to the
Euler angle $\alpha$ renders the direct tunneling to zero. Interestringly, given an equal spacing discretization, i.e., $\alpha_l = 2\pi l /N$ for $l=0, \ldots N-1$, the discretized group averaging
is exactly zero, because of $\sum_{l=0}^{N-1} e^{2\pi i l/N} =0$ for $N>2$. This exact canceling of
the phases points towards very efficient dynamical decoupling protocols in case the perturbation is known.




\emph{Conclusion and outlook.}
The core of the presented quantum control approach to gauge invariance 
relies on the ability to realize the Gauss's law operators. Hence, one can envision 
different gauge groups, as well as different matter content. 
Moreover, different realizations of the matter and gauge degrees of freedom are 
possible as well such as multiple bosons, multiple fermions, or extended objects, like rigid bodies. 
Although we discussed only the Kogut-Susskind Hamiltonian, other formulations of lattice gauge theories such as quantum link models~\cite{Horn1981, Orland1990, Chandrasekharan1997}, or prepotential formulations~\cite{Mathur2005, Anishetty2009}, become feasible within the dynamical decoupling approach. The efficiency of the Haar measure sampling relies on the local nature of $H_0$,
hence the proposed driving scheme remains possible for $d>1$. In the above discussion we used a periodic Cartesian sampling of the Haar measure, however, alternatively one could choose different coordinates or grids, use random unitary sampling~\cite{Elben2018}, or space filling curves~\cite{Bader2012}, e.g. Hilbert curves, which would lead to a periodic continuous protocol. 
While we demonstrated above that quantum control techniques, in particular, dynamical decoupling, can be used to enforce gauge invariance, other quantum control schemes, such as closed loop, or measurement based quantum control schemes can be more efficient. {These closed loop approaches can also be used to enforce global symmetries in quantum many-body systems, for open loop approaches see~\cite{Else2020,Agarwal2020}}. 
In a next step the above dynamical decoupling
approach should be investigated for concrete implementations with artificial quantum system, where prominent candidates are ultracold atom systems, trapped ions or superconducting qubits, which all provide the necessary vertex and link structure for lattice gauge theories.  

\emph{Acknowledgements.}
We thank L.~Barbiero, J.~Berges, D.~González-Cuadra, J.C.~Halimeh, P.~Hauke, and P.~Zoller for
fruitful discussions.\!
ICFO\! group acknowledges support from ERC AdG NOQIA, Spanish Ministry of Economy and Competitiveness (“Severo Ochoa”\!program for Centres of Excellence in R\&D (CEX2019-000910-S), Plan National FIDEUA PID2019-106901GB-I00/10.13039/501100011033, FPI), Fundació Privada Cellex, Fundació Mir-Puig, and from Generalitat de Catalunya (AGAUR Grant No. 2017 SGR 1341, CERCA program, QuantumCAT  \textunderscore U16-011424 co-funded by ERDF Operational Program of Catalonia 2014-2020), MINECO-EU QUANTERA MAQS (funded by State Research Agency (AEI) PCI2019-111828-2/10.13039/501100011033), EU Horizon 2020 FET-OPEN OPTOLogic (Grant No 899794), and the National Science Centre, Poland-Symfonia Grant No. 2016/20/W/ST4/00314. This project has received funding from the European Union's Horizon 2020 research and innovation programme under the Marie Skłodowska-Curie grant agreement No. 754510. This work was supported by the Simons Collaboration on UltraQuantum Matter, which is a grant from the Simons Foundation (651440, P.Z.).
F. J. 
acknowledges the DFG support through the project FOR
2724, the Emmy- Noether grant (Project-ID 377616843).
This work is part of and supported by the DFG Collaborative Research Centre “SFB 1225 (ISOQUANT).
E.Z. acknowledges the support of the ISRAEL SCIENCE FOUNDATION
(grant No. 523/20).

\bibliography{ref}

\end{document}


\section{Supplemental material}

\subsection{Group averaging}

For a  finite group $G$ one defines the group averaging $\Pi(\cdot)$ of an operator $\mathcal{O}$ as
\begin{equation}
\Pi(\mathcal{O})= \frac{1}{|G|} \sum_{g_{j} \in G} V^{\dagger} (g_{j}){\mathcal{O}} V(g_{j}) \, ,
\end{equation}
where $V(g)$ is a unitary operator forming a representation of $G$ parametrized by the group element $g$ and the number of group elements is $|G|$. From the rearrangement lemma \cite{Tung1985} we conclude 
$V^{\dagger}(g_{j})\Pi(\mathcal{O}) V(g_{j}) = \Pi(\mathcal{O})$ for all $g_{j}$, i.e., $\Pi(\cdot)$ symmetrizes the operator $\mathcal{O}$. 
The group averaging can be generalized to continuous groups assuming the existence of a Haar 
measure $d\mu(g)$. We define the group averaging for compact Lie groups, such as SU(N), as
\begin{align}
    \Pi(\mathcal{O}) = \frac{1}{\text{Vol}(G)} \int d\mu(g) V^{\dagger}(g) \mathcal{O} V(g) \, ,
\end{align}
where $V(g)$ is again a unitary representation of $G$. 
The group averaging renders $\Pi(\mathcal{O})$ invariant with respect to group transformations as can be seen from 
\begin{align}
   V^{\dagger}(g_0) \Pi(\mathcal{O}) V(g_0)= \frac{1}{\text{Vol}(G)} \int d\mu(g) V^{\dagger}(g g_0) \mathcal{O} V(g g_0) = \Pi(\mathcal{O}) \, ,
\end{align}
where we used $V(g)V(g_0) = V(gg_0)$ with $g_0$ arbitrary and the invariance of the 
Haar measure $d\mu(g' g_0) = d\mu(g')$ for all $g_0$.
The discretization of the Haar integral shows the connection to the finite case
\begin{align}
    \Pi(\mathcal{O}) \approx \frac{1}{\text{Vol}(G)} \sum_{j} \Delta g_j V^{\dagger}(g_j) \mathcal{O} V(g_j) \, , \label{eq:DiscretizedHaarMeasure}
\end{align}
where we introduced a discrete volume element $\Delta g_j$ and the number $j$ labels the evaluation points. 

\subsection{Hilbert space, algebra and matrix elements on links for SU(2)}
The Hilbert space for SU(2) links is spanned by vectors $\ket{g} \equiv \ket{\alpha, \beta, \gamma}$, 
where $\alpha$, $\beta$, $\gamma$ are the Euler angles of a SU(2) group transformation. The definition
of the link variables is $U_{m n}^{j}$ is given by
\begin{align}
U_{m n}^{j}=\int d\mu(g) D_{m n}^{j}(g)|g\rangle\langle g|\, ,
\end{align}
where $\int d\mu(g)$ is the Haar measure and  $D_{m n}^{j}(g)$ is the $j$-th irreducible representation 
of SU(2). When considering the SU(2) transformations on the link Hilbert space one obtains the SU(2) Lie-algebra. For our purposes we focus on the $j=1/2$ representation and obtain
\begin{subequations}
\begin{align}
{\left[R_{a}, R_{b}\right]=i \varepsilon_{a b c} R_{c}} \, , 
{\left[L_{a}, L_{b}\right]=-i \varepsilon_{a b c} L_{c}} \, ,
{\left[L_{a}, R_{b}\right]=0} \, ,\\
{\left[R_{a}, U_{m n}\right]=U_{m n^{\prime}}\left(\tfrac{1}{2}\sigma_{a}\right)_{n^{\prime} n}} \, ,
{\left[L_{a}, U_{m n}\right]=\left(\tfrac{1}{2} \sigma_{a}\right)_{m m^{\prime}} U_{m^{\prime} n}} \, ,
\end{align} \label{eq:Algebra}
\end{subequations}
with $a,b = 1,2,3$, $\sigma_a$ the Pauli matrices, and the Levi-Civita tensor $\varepsilon_{abc}$.
The Casimir operator is defined by
\begin{align}
\mathbf{J}^{2} \equiv \sum_{a=1}^3 R_{a} R_{a}= \sum_{a=1}^3 L_{a} L_{a} \,.
\end{align}
Because of the commutation relations Eq.\eqref{eq:Algebra}, the operators $L_3$, $R_3$ and $\mathbf{J}^{2}$ form a complete set of commuting operators with the basis vectors $\ket{jmn}$ fulfilling
\begin{subequations}
\begin{align}
\mathbf{J}^{2}  \ket{jmn} &= j(j+1)   \ket{jmn}\, ,  \\
    L_3 \ket{jmn} &= m  \ket{jmn} \, ,\\
    R_3 \ket{jmn} &= n  \ket{jmn} \,.
\end{align}
\end{subequations}
The matrix element of the $U_{mn}$ operators are 
\begin{align}
\left\langle KNN' | U_{mn} | JMM'\right\rangle =
 \sqrt{\frac{2J+1}{2K+1}}
\left\langle JM \tfrac{1}{2} m | KN \right\rangle\left\langle KN' | JM' \tfrac{1}{2} n \right\rangle  \, ,
\end{align}
where the vectors $\left\langle JM \tfrac{1}{2} m | KN \right\rangle$ are Clebsh-Gordon coefficients. For a detailed discussion of the matrix elements we refer to~\cite{Zohar2015b}.



\bibliography{ref}